\documentclass[a4paper,conference]{IEEEtran}
\IEEEoverridecommandlockouts
\usepackage[sort,nocompress]{cite}
\usepackage{hyperref}
\usepackage{amsmath,amssymb,amsfonts}
\usepackage{algorithmic}
\usepackage{graphicx}
\usepackage{textcomp}
\usepackage{xcolor}
\usepackage{array}
\usepackage{tabularx}
\usepackage{booktabs}
\def\BibTeX{{\rm B\kern-.05em{\sc i\kern-.025em b}\kern-.08em
    T\kern-.1667em\lower.7ex\hbox{E}\kern-.125emX}}
\begin{document}

\linespread{0.95}

\title{Verified DRAM Address Mapping Reverse Engineering with Threshold Sweep}

\author{\IEEEauthorblockN{Hewen Zhang}
\IEEEauthorblockA{\textit{University of California, Irvine} \\
Irvine, USA \\
hewenz1@uci.edu}
\and
\IEEEauthorblockN{Ian G Harris}
\IEEEauthorblockA{\textit{University of California, Irvine} \\
Irvine, USA \\
iharris@uci.edu}

}

\maketitle

\begin{abstract}
We propose an end-to-end software pipeline for solving and validating DRAM address mapping functions on embedded LPDDR5 systems, and we instantiate it on embedded platforms. Starting from a large latency dataset of address pairs, the pipeline first builds a latency histogram and then performs a threshold sweep to obtain candidate same-bank address pairs. For each threshold, we apply a Knock-Knock-style \cite{plin2025knock} linear-constraint solver to recover candidate bank masks. Final bank masks are calculated using $\mathrm{GF}(2)$ reduced row echelon form (RREF). To verify correctness, we implement two verification stages, group verification and single function verification, inspired by prior work. Group verification checks whether bank partitions induced by the recovered masks agree with observed timing behavior. Single function verification removes one addressing function at a time and measures the resulting change in verification pass rate. Unlike approaches that return only a single inferred mapping, our workflow systematically records, for each candidate threshold, the solved bank masks and validation metrics. This pipeline supports more robust selection of bank masks on ARM-based embedded devices and provides an efficient method for reverse engineering DRAM address mapping functions and subsequent Rowhammer analysis.
\end{abstract}


\section{Introduction}
Dynamic random-access memory (DRAM) is a core memory technology used by modern computing systems. Kim et al.\cite{kim2014flipping} first systematically revealed DRAM disturbance errors in commodity devices, showing that repeatedly activating a DRAM row can induce bit flips in adjacent rows, which makes Rowhammer a serious security threat. Subsequent work has demonstrated Rowhammer attacks across a wide range of platforms and threat models, including commodity desktops, cloud and cross-VM settings, mobile ARM devices, browsers via JavaScript, and newer CPU ecosystems such as AMD Zen and RISC-V \cite{jattke2025mcsee,jattke2022blacksmith,van2016drammer,jattke2024zenhammer,xiao2016one}. 

To efficiently trigger Rowhammer bit flips, it is important to understand the DRAM address mapping function, which determines how physical addresses are mapped to DRAM locations. These mapping functions enable an attacker to select addresses required for effective hammering, and prior work has shown that better recovered mappings can substantially improve the success of double-sided Rowhammer experiments \cite{pessl2016drama,wang2020dramdig,wi2025sudoku}. However, these functions are typically undocumented and vendor-specific, forcing researchers to reverse engineer them from externally observable behavior \cite{pessl2016drama,xiao2016one}. 
A recent work is Knock-Knock, which departs from the exhaustive search by formulating bank masks recovery in linear-algebraic terms and replacing mask enumeration with a nullspace-based algorithm\cite{plin2025knock}. 

Despite this progress, there are still some limitations. Prior work still focuses mainly on desktop and server-class Intel/AMD systems, while fewer works target embedded platforms especially ARM-based systems with LPDDR5 memory. Moreover, many works report recovered DRAM address mapping functions, but systematic software-only verification of whether the functions are complete and correct is limited. 
Recent work highlights this issue and introduces systematic methods to reliably verify DRAM address mapping functions and individual function components, which provides a more rigorous foundation for evaluating the reverse engineering methods \cite{heckel2025verifying}.

This paper makes the following contributions:
\begin{itemize}
    \item To the best of our knowledge, this is the first work to systematically study DRAM address mapping reverse engineering and verification on an embedded LPDDR5 platform in the context of Rowhammer analysis.

    \item We propose a software-only end-to-end automatic pipeline involving latency measurements, reverse engineering, and verification processes without prior knowledge of DRAM hierarchy.

    \item We instantiate the pipeline on two ARM-based embedded platforms with different memory generations, Jetson Orin Nano with LPDDR5 memory and Raspberry Pi 4B with LPDDR4 memory, showing that the approach is effective beyond a single device or DRAM configuration.


    \item Our pipeline recovers DRAM address mapping functions that are strongly supported by verification. On both platforms, the final solutions achieve pass rates close to 100\% in group verification, and on Jetson Orin Nano the recovered masks are further supported by single function verification results that closely match the expected behavior.

\end{itemize}

\section{Background}

\subsection{DRAM Organization}

Modern DRAM systems are organized hierarchically. At the system level, a memory controller maps physical addresses to a tuple of DRAM components, typically including channel, rank, bank group, bank, row, and column \cite{jedec_lpddr5_jesd209_5c}. 
Each DRAM bank contains a row buffer that temporarily stores the currently activated row. When a row is opened by an \emph{ACTIVATE} command, the entire row is copied into the row buffer, and subsequent \emph{READ} or \emph{WRITE} operations access data from this buffered row at column granularity. Accessing a different row in the same bank requires closing the current row with \emph{PRECHARGE} command and activating the new row \cite{kim2014flipping}. 
This organization creates a timing side channel. If two addresses map to the same bank but different rows, repeatedly accessing them sequentially causes row-buffer conflicts, since each access must evict the previously opened row and reopen the other one, leading to higher latency \cite{pessl2016drama}. In contrast, accesses to different banks do not have any conflicts due to bank parallelism. 
By measuring these latency differences, same bank address pairs can be distinguished from different bank address pairs, which forms the basis of many DRAM address mapping reverse engineering techniques \cite{pessl2016drama,wang2020dramdig}.

LPDDR5 complicates this problem further because it supports multiple bank modes, including BG mode, 8B mode, and 16B mode \cite{jedec_lpddr5_jesd209_5c}. As a result, even knowing a part number of the DRAM chip does not necessarily reveal the number of banks in a target system. However, many prior works \cite{jattke2024zenhammer,pessl2016drama,wang2020dramdig} need the number of banks as input. Consequently, these methods are less suitable for modern LPDDR5 systems and DRAM-agnostic approaches are needed.

\subsection{DRAM Addressing}

The memory controller implements a set of DRAM address mapping functions that map a physical address to a DRAM tuple, e.g., channel, rank, bank group, bank, row, and column. Prior work has shown that many of these functions can be modeled as linear functions over $\mathrm{GF}(2)$, i.e., XOR combinations of physical address bits \cite{pessl2016drama,plin2025knock}. Under this view, each output bit of an addressing function is represented by a binary bank mask over physical address bits. Evaluating the bank mask on an address means XORing the selected bits. In this paper, we use the term \emph{bank mask} to denote one such XOR-based function component.

Traditional software-based reverse engineering often recovers DRAM address mapping functions by exhaustive search. DRAMA, and many subsequent methods derived from the same general idea, measure row-buffer conflicts and then enumerate candidate XOR functions to find the set that best explains the observed conflict relation \cite{pessl2016drama,wang2020dramdig,jattke2024zenhammer,jattke2025mcsee,kaur2023flipping}. This approach can work well, but the search space grows rapidly with the number of candidate bits and independent mapping functions. It also tends to require platform-specific knowledge to remain tractable on modern systems \cite{wang2020dramdig,plin2025knock}.

Knock-Knock replaces this exhaustive-search paradigm with a linear-algebraic formulation \cite{plin2025knock}. Its key idea is that if two physical addresses map to the same bank, then every addressing function must have the same output on both addresses. Let $a$ and $b$ be two same-bank addresses, and let $\Delta = a \oplus b$. Equivalently, $a \oplus b \oplus \Delta = 0$. Since each bank mask $m$ is modeled as a linear parity function over $\mathrm{GF}(2)$, the equality of bank outputs implies
\[
\begin{array}{rcl}
m(a) &=& m(b) \\
&\Leftrightarrow& m(a)\oplus m(b)=0 \\
&\Leftrightarrow& m(a\oplus b)=m(\Delta)=0
\end{array}
\]
Therefore, for any valid bank mask $m$, we obtain
$
m \cdot \Delta = 0 \pmod 2,
$
where the dot product is taken over $\mathrm{GF}(2)$. Thus, every empirically identified same bank pair induces one homogeneous linear constraint on the unknown bank masks. By stacking many such constraints into a difference matrix $D$, the valid bank masks must lie in the nullspace of $D$ over $\mathrm{GF}(2)$:
$
D \cdot m^{T} = 0.
$
Instead of enumerating all candidate functions, Knock-Knock therefore computes a basis of this nullspace and interprets it as the recovered bank mask basis.

Using this approach, the problem is no longer an exponential search over all XOR combinations, but a polynomial-time linear algebra on constraints extracted from timing measurements. In addition, the method is black-box and platform-agnostic. 

\section{Methods}\label{methods}

Our approach has two steps. The first step, \textbf{Latency measurement}, allocates a fraction of the whole memory and repeatedly measures the latency of sequentially accessing two randomly selected addresses, thereby collecting a large number of latency samples. The second step, \textbf{Threshold sweep}, iterates a possible range of latencies and performs same-bank pairs extraction, bank mask computation, group verification and single function verification. 

The details of \textbf{Threshold sweep} are described below.
\begin{itemize}
    \item \textbf{Histogram generation}: A histogram of all measured latencies and a list of possible thresholds is generated.
    \item \textbf{Same-bank pairs extraction}: Our approach iterates a range of possible thresholds and for each threshold, same-bank pairs are extracted based on it. If a pair of addresses reside in different rows in the same bank, the latency of accessing them sequentially is higher than the latency of accessing other types of address pairs. As a result, those address pairs with higher latencies than the threshold will be considered to be same bank pairs.
    \item \textbf{Bank mask computation}: Same-bank pairs are used by the Knock-Knock algorithm \cite{plin2025knock} to compute the bank masks. Same-bank pairs are randomly sampled from the candidate set. For each sampled pair, the XOR of the two addresses is computed, and the resulting XOR vectors are stacked row by row to form a difference matrix. Then, the nullspace of the difference matrix is computed, and Gaussian elimination over $\mathrm{GF}(2)$ is applied to the resulting nullspace basis vectors to obtain their reduced row echelon form (RREF). The resulting RREF basis gives the final bank masks.
    \item \textbf{Group verification} and \textbf{Single function verification}: Our approach verifies the bank masks based on the algorithm proposed in \cite{heckel2025verifying}. Verification is needed because the recovered bank masks are inferred results, and whether they are consistent with the observed latency behavior must be checked. Group verification evaluates the correctness of the bank mask set as a whole, while Single function verification examines the correctness of each individual bank mask. For \textbf{Group verification}, $20{,}000$ addresses from an allocated memory region are placed into an address pool. The bank ID of each address in the pool are computed and all the addresses are partitioned into $2^{\#\text{bank masks}}$ groups. In each group, an address $A_1$ is randomly selected. Another address $A_2$ from the same bank and an address $B_1$ from a different bank are then randomly chosen. The latency $t_1$ for accessing $A_1$ and $A_2$ is measured, and the latency $t_2$ for accessing $A_1$ and $B_1$ is also measured. If $t_1 > \text{threshold}$ and $t_2 < \text{threshold}$, this iteration is counted as a pass; otherwise, it is counted as a failure. After repeating this process for many addresses and across all banks, the pass rate is computed to represent the reliability of the recovered bank masks. For \textbf{Single function verification}, the main purpose is to verify the correctness of each bank mask in the recovered bank mask set. If one correct bank mask is removed from the set, the pass rate is expected to be reduced to approximately half of that of \textbf{Group verification}. Each bank mask is removed in turn and the pass rate of the remaining bank masks will be measured.

\end{itemize}



\begin{table*}[!t]
\caption{Recovered bank masks on Jetson Orin Nano and Raspberry Pi 4B}
\label{tab:bankmasks}
\centering
\footnotesize
\setlength{\tabcolsep}{4pt}
\renewcommand{\arraystretch}{1.1}
\begin{tabularx}{\textwidth}{|>{\centering\arraybackslash}m{2.7cm}
                            |>{\centering\arraybackslash}m{1.5cm}
                            |X
                            |>{\centering\arraybackslash}m{2.0cm}|}
\hline
\textbf{Platform} & \textbf{\# Masks} & \textbf{Recovered bank masks} \\
\hline
Jetson Orin Nano & 7 & 0x214ccde00, 0x110ad4400, 0x950ccc00, 0x50eeac00, 0x21708c00, 0xd331200, 0x235d600 \\
\hline
Raspberry Pi 4B & 4 & 0x80000000, 0x4000, 0x2000, 0x1000 \\
\hline
\end{tabularx}
\end{table*}


\section{Results}

\subsection{Recovered Bank Masks and Verification Results} \label{recover}

Table~\ref{tab:bankmasks} summarizes the final bank masks recovered on Jetson Orin Nano and Raspberry Pi 4B using the pipeline described in Section~\ref{methods}. Our approach collects $10{,}000{,}000$ latencies in \textbf{Latency measurement} and the latency represents CPU clock cycles. For \textbf{Bank mask computation}, the computation is conducted $200$ times and each time, $100$ different same-bank pairs are subsampled to form a difference matrix. Each row of the difference matrix is the XOR vectors of the two same-bank addresses.

\begin{figure}[htbp]
\centering
\includegraphics[width=\linewidth]{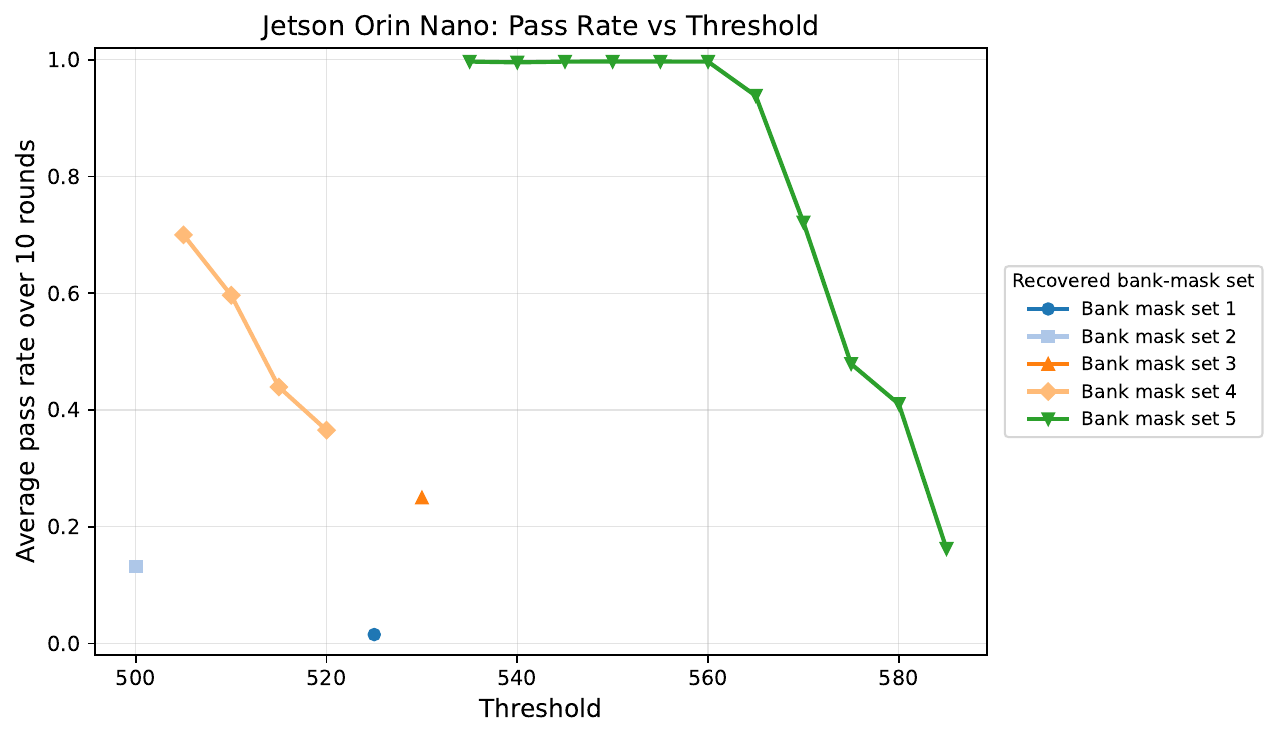}
\caption{\textbf{Group verification} pass rate versus threshold on Jetson Orin Nano. Each curve corresponds to one recovered bank mask set, and different colors/markers denote different solutions.}
\label{fig:group1}
\end{figure}

Figure~\ref{fig:group1} report the \textbf{Group verification} pass rate for each threshold as a function of threshold for Jetson Orin Nano. For several bank mask sets, only a single threshold produces a result, so it appears a single point in Figure~\ref{fig:group1}. The pass rate in the figure is the average value of the pass rates after executing  \textbf{Group Verification} for 10 rounds. For Jetson Orin Nano, \textbf{Threshold sweep} produces five distinct recovered bank mask sets. 

Among these candidates, bank mask set~5 consistently yields the highest group verification pass rate and reaches a region close to $1.0$ over a broad threshold range. In contrast, the lower-dimensional candidate sets achieve substantially lower pass rates. This separation suggests that, on Jetson Orin Nano, bank mask set~5 is the candidate most consistent with the observed same-bank versus different-bank latency behavior.

A similar trend is observed on Raspberry Pi 4B. The \textbf{Threshold sweep} yields two candidate bank mask sets: bank mask set~1 contains \texttt{0x80000000, 0x2000, 0x1000}, bank mask set~2 contains \texttt{0x80000000, 0x4000, 0x2000, 0x1000}. Bank mask set~2, dominates the high pass rate region and remains close to $1.0$ across most of the selected threshold range, while bank mask set~1 yields negligible pass rate. Consequently, the \textbf{Group verification} results on both platforms show that the recovered bank masks produce bank partitions that agree  with the measured latency behavior. Accordingly, the final bank masks reported in Table~\ref{tab:bankmasks} are the solutions most strongly supported by the verification data.

We further evaluate the bank masks of Jetson Orin Nano using \textbf{Single function verification}. The selected threshold is $550$, where the pass rate of \textbf{Group verification} is $0.996907$, the highest among the threshold range. The expected pass rate after removing one correct bank mask is $0.498454$. For each of the bank masks, removing that mask yields a pass rate very close to the expectation. This behavior is consistent with the principle of \textbf{Single function verification}, which means each recovered addressing function contributes independently to the final bank partition, and removing any one of them causes a degradation in verification quality. This result provides additional evidence that the recovered bank masks on Jetson Orin Nano are stable and reliable.

\subsection{Threshold Sensitivity and Verification Analysis}

\begin{figure}[htbp]
\centering
\includegraphics[width=\linewidth]{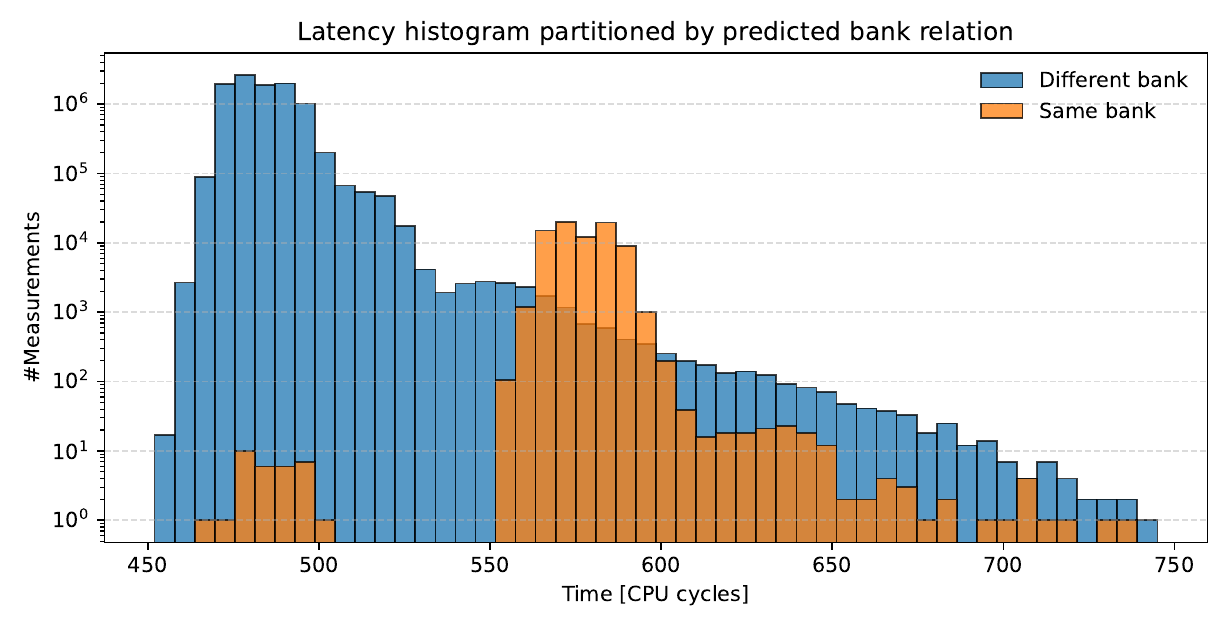}
\caption{Latency histogram partitioned by predicted bank relation.}
\label{fig:hist}
\end{figure}



A more detailed analysis on Jetson Orin Nano is performed to better understand threshold sensitivity. This analysis is based on the latency data collected during \textbf{Latency measurement} and examines the recovered bank masks from three perspectives: the latency distribution induced by the recovered partition, the expected \textbf{Group verification} pass rate as a function of threshold, and the contamination behavior under subsampling.

Figure~\ref{fig:hist} shows the latency histogram partitioned by the bank relation predicted by the final recovered bank masks in Table~\ref{tab:bankmasks}. Using the recovered addressing functions, each measured address pair is classified as either \emph{same bank} or \emph{different bank}. The figure shows a clear separation, that is, most different-bank measurements concentrate in the lower-latency region, while the same-bank measurements are in higher latencies. The observation is consistent with the \textbf{Group verification} results in Section~\ref{recover}, which shows that thresholds located in the gap between the two distributions are expected to better separate same-bank from different-bank pairs.

As $10{,}000{,}000$ latency samples are collected and the sample size is large, the resulting empirical distributions are used as a practical approximation of the memory behavior on Jetson Orin Nano. Based on these latency samples and the recovered bank masks, the expected \textbf{Group verification} pass rate is estimated at each threshold. 
Let $L_{\mathrm{same}}$ and $L_{\mathrm{diff}}$ denote the latency random variables for pairs predicted to be in the same bank and in different banks, respectively. For a threshold $\tau$, define
\[
p_{\mathrm{same}}(\tau)=\Pr[L_{\mathrm{same}}>\tau], \qquad
p_{\mathrm{diff}}(\tau)=\Pr[L_{\mathrm{diff}}<\tau].
\]
Under the natural independence approximation between the two sampled events in one verification trial, the expected pass probability is
$
p_{\mathrm{pass}}(\tau)=p_{\mathrm{same}}(\tau)\,p_{\mathrm{diff}}(\tau).
$
These probabilities are estimated empirically from the latency dataset obtained during \textbf{Latency measurement}. 
The expected pass rate is close to zero at very low and very high thresholds, and reaches a region near $1.0$ in an intermediate threshold range. This follows from the two histogram tails. If $\tau$ is too small, many different-bank pairs also exceed the threshold. If $\tau$ is too large, many same-bank pairs fail to exceed it. Only thresholds near the separation region, approx. 500-570 clock cycles, can yield higher verification pass rate. The high pass rate range is therefore consistent with the threshold region in which the final bank masks achieve the highest measured \textbf{Group verification} performance. 

\section{Conclusion}

This paper presented an end-to-end software-only pipeline for reverse engineering and verifying DRAM address mapping functions on embedded systems, with a particular focus on modern LPDDR5 platforms. Our workflow combines latency measurement, threshold sweep, Knock-Knock-based bank mask recovery, and two verification stages, including group verification and single function verification. Unlike prior approaches that return only a single inferred solution, our method records the recovered bank masks and their verification metrics for each candidate threshold. Overall, our results show that combining reverse engineering with rigorous software-based verification provides a practical and robust methodology for DRAM analysis on embedded ARM platforms especially LPDDR5 systems, where DRAM hierarchy information is incomplete.

\bibliographystyle{plain}
\bibliography{ref}

\end{document}